\documentclass[prl,aps,twocolumn,superscriptaddress,showpacs]{revtex4}

\usepackage{epsfig}
\usepackage{amsmath}
\usepackage{graphicx}

\newcommand{\be}{\begin{equation}}
\newcommand{\ee}{\end{equation}}
\newcommand{\bea}{\begin{eqnarray}}
\newcommand{\eea}{\end{eqnarray}}

\newcommand{\p}{\partial}

\newcommand{\la}{\langle}
\newcommand{\ra}{\rangle}

\newcommand{\ham}{{\cal H}}

\def\nn{\nonumber\\}

\begin{document}

\title{The effect of a local perturbation in a fermionic ladder}

\author{Sam T. Carr} \affiliation{Institut f\"ur Theorie der
  Kondensierten Materie and DFG Center for Functional Nanostructures,
  Karlsruher Institut f\"ur Technologie, 76128 Karlsruhe, Germany}
\author{Boris N. Narozhny} \affiliation{Institut f\"ur Theorie der
  Kondensierten Materie and DFG Center for Functional Nanostructures,
  Karlsruher Institut f\"ur Technologie, 76128 Karlsruhe, Germany}
\author{Alexander A. Nersesyan} \affiliation{The Abdus Salam
  International Centre for Theoretical Physics, 34100, Trieste, Italy}
\affiliation{The Andronikashvili Institute of Physics, 0177, Tbilisi,
  Georgia}\affiliation{Ilia State University, Institute of Theoretical
  Physics, 0162, Tbilisi, Georgia}

\date{\today}

\begin{abstract}
  We study the effect of a local external potential on a system of two
  parallel spin-polarized nanowires placed close to each other. For
  single channel nanowires with repulsive interaction we find that
  transport properties of the system are highly sensitive to the transverse
  gradient of the perturbation: the asymmetric part
  completely reflects the electrons leading to vanishing
  conductance at zero temperature, while the flat potential remains
  transparent. We envisage a possible application of this unusual
  property in the sensitive measurement of local potential field
  gradients.
\end{abstract}

\pacs{73.63.Nm, 71.10.Pm, 85.35.Be}

\maketitle

Recent advances in nanotechnology have led to an explosive growth of
experimental work on low-dimensional systems. Single-channel
\cite{ya1} and multi-channel \cite{son,mon} one-dimensional (1D)
conductors or nanowires can now be manufactured in a controlled
fashion. At room temperature, nanowire arrays \cite{ma1,man} have
found real-life applications as highly sensitive chemical \cite{chs},
biological \cite{bhs}, and optical \cite{ops} sensors. Individual
nanowires have been used as mechanical force and mass sensors
\cite{fms}. Furthermore, at low temperatures both nanowires \cite{ya1}
and carbon nanotubes \cite{cnt} provide a fertile ground for
laboratory experiments aimed at uncovering the quantum nature of
interacting many-particle systems \cite{rex}.

The quantum physics of an individual nanowire is well understood
within the paradigm of the Tomonaga-Luttinger (TL) liquid
\cite{llh}. As shown in the seminal paper of Kane and Fisher
\cite{kaf}, transport properties of the TL liquid are very sensitive
to the application of a local external potential. While the clean TL
liquid is a perfect conductor, even a weak impurity potential in the
most physical case of repulsive electron-electron interaction
completely reflects the electrons leading to zero conductance at
temperature $T=0$.

The dramatic response of an interacting 1D system to a local probe is
in sharp contrast with the role of such a probe in higher dimensions.
A natural question to ask then is how a single impurity would affect
an array of nanowires in a situation intermediate between purely one-
and two-dimensional cases.

\begin{figure}
\begin{center}
\epsfig{file=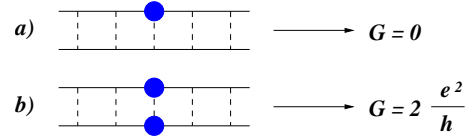,width=6cm}
\end{center}
\caption{[Color online] $T=0$ conductance $G$ of the repulsive $U,V>0$
  regime of model (\ref{ham-kin}) in cases of (a) single impurity and
  (b) two identical impurities placed on a rung.}
\label{lim}
\end{figure}

Consider two repulsive spinless TL liquids away from $1/2$ filling
brought together to form a ladder, as shown in Fig.~\ref{lim}. Assume
that the fermions belonging to different wires interact with each
other and may also undergo inter-chain hopping. Imagine that the
external potential affects a single rung of the ladder and compare the
two situations shown in Fig.~\ref{lim}: (a) a single impurity located
on one of the chains and (b) a pair of identical impurities on the
rung. Naively one might expect that the case (b) could be described by
the Kane-Fisher scenario leading to vanishing conductance at $T\to 0$,
whereas the geometry of the ladder in the case (a) would suggest
ballistic conductance in the zero-temperature limit. As shown in the
present Letter, this conclusion cannot be correct.  Moreover, for the
most physical case of repulsive interaction in the ladder the right
conclusion is just the opposite, with the roles of cases (a) and (b)
interchanged.

Indeed, the Kane-Fisher effect -- the growth of the local potential
upon decreasing the energy scale, or in other words the relevance of
the local perturbation in the renormalization group (RG) sense -- only
occurs if the impurity potential couples to the operator whose
space-time correlations are dominant in the clean system. In the case
of a single {\em repulsive} ($V>0$) TL liquid these are the ``charge
density wave'' (CDW) correlations that locally couple to the
backscattering part of the impurity potential and get pinned at the
impurity site driving the conductance to zero. In a ladder, the
relative phase of the two incommensurate CDWs on individual chains is
$0$ for attractive inter-chain interaction ($U<0$) or $\pi$ for
repulsive inter-chain interaction ($U>0$), the corresponding
density-wave states being labeled by CDW$^{\pm}$. In case (b) the
symmetric potential couples only to the CDW$^+$ operator. Thus, such
an impurity is transparent (at $T=0$) for the repulsive ($U,V>0$)
ladder but suppresses conductance for $U<0$. On the contrary, the
local impurity in case (a) couples to both CDW$^+$ and CDW$^-$ leading
to $G\rightarrow 0$ as $T\rightarrow 0$ for any sign of $U$
\cite{fn1}.

The above argument (and the calculation below) show that the repulsive
two-leg ladder is extremely sensitive not only to the presence of the
local probe, but also to the transverse gradient of the external
potential. Since the ladder is a prototypical model of two interacting
and closely located nanowires, our results suggest a possible use of
double nanowire devices \cite{ya1} as quantum sensors of local field
gradients in low-temperature experiments.

{\em Model. ---} 
Our model is described by the Hamiltonian
\bea
&& H = -\frac{t_\|}{2}\sum_{i,\sigma} \left(
  c_{\sigma,i}^\dagger c_{\sigma,i+1} + H.c.\right) + U\sum_i n_{1,i}
n_{2,i}
\nonumber \\
&&
\nonumber\\
&& - t_\perp\sum_{i}\left( c_{1,i}^\dagger c_{2,i}+H.c.\right) + V
\sum_{i,\sigma} n_{\sigma,i} n_{\sigma,i+1}, 
\label{ham-kin} 
\eea 
\noindent
where $c^\dagger_{\sigma,i}$ is the creation operator for a fermion
residing on site $i$ of the chain $\sigma=1,2$,
$n_{\sigma,i}=c_{\sigma,i}^\dagger c_{\sigma,i}$ is the density
operator, $t_\|$ and $t_\perp$ describe the intra- and inter-chain
hopping, and $V$ and $U$ are the intra- and inter-chain
nearest-neighbor interaction constants. Note that in the absence of
$t_\perp$ and $V$, this is simply the Hubbard model where the two legs
of the ladder can be mapped to spin-up and spin-down.

For $U,V>0$ this model describes two closely located nanowires coupled
by the Coulomb interaction \cite{ya1}. If the wires are sufficiently
far apart (e.g. for inter-wire distance of order $50 nm$ \cite{man}),
 $t_\perp$ may be set to zero (we show below this is unimportant for our purposes).
 


A local (at $i=0$) external potential is described by 
\be
H_{\text{imp}} = \sum_\sigma \lambda_\sigma n_{\sigma,i=0}, 
\label{imp}
\ee 
\noindent
where $\lambda_{1,2}$ is the impurity strength on the top and bottom
chain respectively. We distinguish the two limiting cases: (a) a
purely local impurity, corresponding to $\lambda_2=0$, and (b) a flat
transverse potential that is equivalent to two identical local
impurities, $\lambda_1=\lambda_2$ (see Fig.~\ref{lim}).

Note that the most general form of the external field affecting the
single rung of the ladder has the form $\lambda^a c^\dagger_\sigma
\tau^a_{\sigma\sigma'} c_{\sigma'}$ (where $\tau^a$ are the Pauli
matrices) which in addition to Eq.~(\ref{imp}) [$\lambda^{0(z)} =
(\lambda_1\pm\lambda_2)/2$] contains a local variation of the
inter-chain hopping ($\lambda^x$) and a local magnetic field
($\lambda^y$) -- see Table~\ref{dop}. As it turns out these terms are
not important for repulsive interaction.

\begin{table}[tbp]
\caption{Local operators in the fermionic and bosonic forms
(backscattering component only). CDW$^{\pm}$ refer to charge-density
waves at a relative phase of $0$ or $\pi$. OAF refers to the orbital
anti-ferromagnet, while BDW is a bond-density wave.}
\begin{ruledtabular}
\begin{tabular}{ccc}
Local operator & Fermionic form & Bosonic form \\
\hline
${\cal O}_{\text{CDW}^+}$ & $c_\sigma^\dagger \tau^0_{\sigma\sigma'}c_{\sigma'}$ 
& $\sin \sqrt{2\pi} \phi_c \cos \sqrt{2\pi} \phi_s$ \\
${\cal O}_{\text{BDW}}$ & $c_\sigma^\dagger \tau^x_{\sigma\sigma'} c_{\sigma'}$ 
& $\cos \sqrt{2\pi} \phi_c \sin \sqrt{2\pi} \phi_s$ \\
${ \cal O}_{\text{OAF}}$ & $c_\sigma^\dagger \tau^y_{\sigma\sigma'} c_{\sigma'}$ 
& $\cos \sqrt{2\pi} \phi_c \cos \sqrt{2\pi} \theta_s$ \\
${\cal O}_{\text{CDW}^-}$ & $c_\sigma^\dagger \tau^z_{\sigma\sigma'} c_{\sigma'}$ 
& $\cos \sqrt{2\pi} \phi_c \sin \sqrt{2\pi} \theta_s$ 
\end{tabular}
\end{ruledtabular}
\label{dop}
\end{table}

{\it Qualitative discussion. ---} 
In the absence of electron-electron interaction the single impurity
problem can be solved exactly for the model (\ref{imp}) or treated
using the scattering matrix approach 
 for arbitrary shape of
the external potential. The impurity induces a modulation of particle
density. Taking into account the interaction one finds an additional
scattering potential due to the oscillating part of the modulation
known as the Friedel oscillation. For a single nanowire \cite{mag}, already
within the Hartree-Fock approximation one can observe the divergence
of the (back-)scattering amplitude providing an alternative
explanation for the Kane-Fisher effect.

Friedel oscillations can also be found in the ladder model.
 It is clear that in the presence of inter-chain hopping
($t_\perp\ne 0$) the single impurity ($\lambda_2=0$) will induce
Friedel oscillations in both chains. What is most important, the
oscillations on the two chains are out of phase. In other words, a
charged impurity placed on one chain of the ladder will induce an
image charge on the other chain which will be of the opposite sign.

However unlike the single-wire Kane-Fisher problem, the argument based
on Friedel oscillations can not be extended further in order to
explain our results. By design, this is a weak-coupling argument.  As
we discuss below, the excitation spectrum of ladder models contains
gapped branches, which can not be accounted for by any weak-coupling
expansion. Moreover, the existence of the Friedel oscillation is
essentially tied to the presence of the inter-chain hopping
\cite{fn2}, which it turns out plays only a minor role in the
longitudinal transport of the repulsive ladder. In what follows, we
therefore focus on the Kane-Fisher idea \cite{kaf} of pinning the
incommensurate density waves by the local impurity, qualitatively
outlined above.

{\em Calculation. ---} 
We now outline the derivation of our results (full details will be
given elsewhere \cite{dopo}). First, we diagonalize the
single-particle problem (without the impurity) and obtain the two-band
spectrum. Second, we linearize the spectrum about the
Fermi points and apply the standard bosonization procedure
\cite{bos} with the conventions \cite{us2} $R(L)_{\mu} =
(\kappa_{\mu}/\sqrt{2\pi\alpha}) \exp(\pm i\sqrt{4\pi}
\phi^{R(L)_{\mu}})$, where $R(L)_{\mu}$ are the operators of right
(left) movers in the band $\mu=\pm$ and $\alpha$ is the bosonic
ultraviolet cut-off.  The chiral bosonic fields commute as $\left[
  \phi^R_{\mu}, \phi^L_{\mu'} \right] = i\delta_{\mu\mu'}/4$
and combine in the usual way $\phi_{\mu} = \phi^L_{\mu} +
\phi^R_{\mu}$ and $\theta_{\mu} = \phi^L_{\mu}-\phi^R_{\mu}$.  The
anti-commuting Klein factors, $\left\{ \kappa_{\mu}, \kappa_{\mu'}
\right\} = 2\delta_{\mu\mu'}$, are not dynamic variables, so we can
choose the representation $\kappa_1\kappa_2 = i$ and $\kappa_\mu^2=1$.
Finally, we arrange the bosonic fields into the charge
$\phi_c=(\phi_++\phi_-)/\sqrt{2}$ and pseudo-spin
$\phi_s=(\phi_+-\phi_-)/\sqrt{2}$ modes (c.f. spin-charge separation).

\begin{figure}
\begin{center}
\includegraphics[width=3.1in,clip=true]{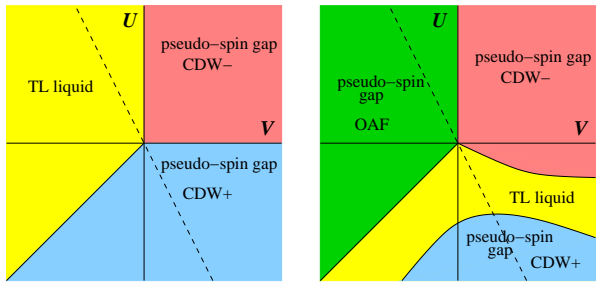}
\end{center}
\caption{[Color online] Phase diagram of model (\ref{ham-kin}) at an
  incommensurate filling. The left panel corresponds to $t_\perp=0$ and
  the right panel to $t_\perp\ne 0$. The dashed line corresponds to
  $K_c=1$. To the right of the dashed line the local density operators
  (see Table~\ref{dop}) exhibit dominant correlations. To the left of
  the dashed line the dominant correlations are super-conducting,
  however the impurity (\ref{imp}) does not couple to such operators
  which are therefore irrelevant for our purposes.}
\label{pd}
\end{figure}

The precise form of the effective low-energy theory depends on the
value of the inter-chain hopping $t_\perp$. Having in mind
experimental realizations \cite{ya1,man} of nanowires with a
sufficient distance between then, we start with the case of
$t_\perp=0$. Then the bosonized form of the Hamiltonian
(\ref{ham-kin}) is equivalent to the XYZ Thirring model \cite{xyz}:
\begin{gather}
\ham_c = \frac{v_F}{2} \left[ K_c \Pi_c(x)^2 + 
\frac{1}{K_c} \left( \p_x \phi_c(x) \right)^2 \right], 
\nn
\ham_s = \frac{v_F}{2} \left[K_s  \Pi_s(x)^2 + 
\frac{1}{K_s} \left( \p_x \phi_s(x) \right)^2 \right] 
\nn
+ \frac{g_\perp}{2(\pi \alpha)^2} \cos \sqrt{8\pi} \phi_s(x) +  
\frac{g_f}{2(\pi \alpha)^2} \cos \sqrt{8\pi} \theta_s(x).
\label{bozband}
\end{gather}
For weak interaction the parameters of the Hamiltonian (\ref{bozband})
can be related to the original interaction constants $U$ and $V$
through ($a$ is the lattice spacing)
\begin{gather}
K_c=1-(2\bar{V}+\bar{U})/\pi v_F, \quad K_s = 1+\bar{U}/\pi v_F, \nn
g_\perp = \bar{U}-\bar{V}, \quad g_f = \bar{V}, \nn
\bar{U}=aU, \quad \bar{V}=aV[1-\cos(2k_Fa)].
\label{KUV}
\end{gather}
\noindent
The phase diagram of the effective model (\ref{bozband}) is shown in
the left panel of Fig.~\ref{pd} (as obtained using the usual one-loop
RG approach \cite{bos,rg}). Note that that in all phases the charge
sector of the model remains gapless and can be described as a TL
liquid.

The effect of $t_\perp$ which adds the term $-t_\perp \sqrt{(2/\pi)}
\partial_x\phi_s$ to the Hamiltonian (\ref{bozband}) is well studied
\cite{NLK,rg,us2}. For repulsive ladders this has little effect on the
ground state of the model (hence our statement of the irrelevance of
the value of $t_\perp$ to the conductance properties of physical
double nanowire devices), see the phase diagram in the right panel of
Fig.~\ref{pd}. The most notable change in the phase diagram is the
appearance of the orbital anti-ferromagnet (OAF) phase for the case of
intra-chain attraction $V<0$ \cite{NLK,oaf}, which is impossible for
$t_\perp=0$.

In the pseudo-spin-gap phases one of the nonlinear terms in the
Hamiltonian (\ref{bozband}) scales to strong coupling and becomes
relevant in the RG sense. Consequently, the corresponding bosonic
field gets locked to the value defined by one of the minima of the
cosine potential depending on the sign of the interaction
parameter. In particular, for the repulsive ladder this scenario is
realized for the last, $g_f$-proportional term, so that the field
$\theta_s$ gets locked to the value $\sqrt{\pi/8}+m\sqrt{\pi/2}$ (with
$m$ being arbitrary integer).  This implies the non-zero expectation
value $\langle\sin\sqrt{2\pi}\theta_s\rangle\ne 0$. Due to the
multiplicative structure of local operators as shown in
Table~\ref{dop} which always include the massless charge field, this
situation does not lead to long-range order. Despite the gap opening
the conductance of the clean ladder \cite{ms} is $G=2e^2/h$
\cite{sta}, as the gapped pseudo-spin mode does not carry charge.

Having now identified the ground state of the clean system, we can
consider what happens when an impurity is added at $i=0$. Applying the
bosonization rules to the local perturbation (\ref{imp}) we find that
the external potential can be written in terms of local operators from
Table~\ref{dop}:
\begin{equation}
\label{bozimp}
H _{\text{imp}} = [(\lambda_1+\lambda_2){\cal O}_{\text{CDW}^+} 
+ (\lambda_1-\lambda_2){\cal O}_{\text{CDW}^-}]/2.
\end{equation}
\noindent
Let us first consider the case of a symmetric impurity
$\lambda_1=\lambda_2$ in the repulsive ladder. Then the perturbation
is proportional to $\cos \sqrt{2\pi} \phi_s$ (see Table~\ref{dop}).
Now $\theta_s$ is locked and, consequently, all correlation functions
of exponents of $\phi_s$ are short-ranged (exponentially decaying). To
obtain the effective theory of the charge sector for energies small
compared to the pseudo-spin gap, we integrate out the pseudo-spin
degree of freedom.  The impurity potential does not contribute in the
first order, while in the second order it generates the pair
backscattering term $\cos\sqrt{8\pi} \phi_c(0)$. For $K_c>1/2$ this
operator is irrelevant is the RG sense, leaving us with a single
channel {\it weak-coupling} Kane-Fisher-type problem. Following
Ref.~\onlinecite{kaf} we see that 
\be 
G = 2 e^2/h - A
\left[\text{Max}(T,\mathcal{V})\right]^{\gamma}, \quad
\gamma=4K_c-2>0,
\label{rb}
\ee 
\noindent
where $\mathcal{V}$ is the voltage bias and $A$ is a non-universal
constant. Thus at $T=0$ the symmetric impurity or, equivalently, the
flat external potential remains transparent.

The situation is substantially different if $\lambda_1\ne \lambda_2$,
i.e. one also has a local perturbation proportional to ${\cal
  O}_{\text{CDW}^-}$. In this case (still for repulsive interaction)
the impurity contributes in the first order as one integrates out the
massive (CDW$^-$ phase) pseudo-spin,  which is
equivalent to replacing $\sin\sqrt{2\pi} \theta_s$ with its
expectation value
\be
\cos \sqrt{2\pi} \phi_c \sin \sqrt{2\pi} \theta_s \rightarrow 
\la \sin \sqrt{2\pi} \theta_s \ra  \cos \sqrt{2\pi} \phi_c.
\label{sco}
\ee
\noindent
We are again left with what looks like a single-channel
Kane-Fisher-type problem for the charge sector, but this time we are
in the {\it strong coupling} regime as the scaling dimension of the
impurity operator is $d=K_c/2$, implying that the operator is relevant
for any $K_c<2$.

In the strong-coupling regime, the single-chain Kane-Fisher problem
reduces to that of a weak link between two semi-infinite TL
liquids. In the ladder model the impurity operator (\ref{sco}) has a
multiplicative structure with its pseudo-spin part frozen due to
ordering of the field $\theta_s$ in the CDW$^-$ phase. Consequently,
in this case the strong-coupling regime is {\it not} equivalent to the
problem of weak tunneling between two semi-infinite ladders, but is
rather determined by the charge sector alone. Henceforth the
calculation is analogous to the single-chain case, except that there
are no single-particle processes left in the effective theory of the
charge channel. One therefore considers pair hopping across the
boundary described by the operator $\cos\sqrt{8\pi}\theta_c$. There is
no conductance at $T=0$, while at $T>0$ the conductance is given by
\be 
G = B \left[\text{Max}(T,{\cal V})\right]^{\gamma'}, 
\quad 
\gamma'=4/K_c - 2 >0.
\label{ra}
\ee 
\noindent
The non-universal coefficient $B$ is proportional to the square of the
effective pair tunneling rate, which is in general unknown.

The two results (\ref{rb}) and (\ref{ra}) seem to be disconnected as
there appears no clear way to obtain one from another by a smooth
variation of the impurity strength. This is not surprising as each of
the two results was obtained by an effective weak-coupling calculation
- a weak impurity in the case of (\ref{rb}) and a weak pair tunneling
in the case of (\ref{ra}). While it might be possible to trace the
connection between the two regimes for a particularly suitable model
\cite{arw}, such calculation would be strongly model-dependent. In
contrast, the form of the results (\ref{rb}) and (\ref{ra}) as well as
the $T=0$ behavior (Fig.~\ref{lim}) are universal.

The rest of the phase diagram shows a similar behavior -- if the
external potential couples to the local operator with dominant
correlations in the bulk, then the impurity strength flows to strong
coupling driving conductance to $G\approx 0$ [see Eq.~(\ref{ra})],
otherwise the conductance remains close to its clean value as in
Eq.~(\ref{rb}). An interesting example is the OAF phase, where the
ballistic conductance may only be suppressed by a local magnetic field
$\lambda^y {\cal O}_{OAF}$. In any of these phases, as temperature is
increased beyond the size of the pseudo-spin gap, the exponents will
smoothly cross over to the spinful TL liquid values given in
Ref.~\onlinecite{kaf}; this also describes transport in the gapless
TL liquid phase.

{\em To summarize ---}
we have considered the effect of a local external potential on
transport properties of a spinless two-leg ladder. We find that the in
the physical case of repulsive interaction the system is extremely
sensitive to the transverse gradient of the potential which grows under renormalization and at $T=0$ completely
suppresses conductance in contrast with the flat potential
that remains transparent.
We note that although any realistic potential will have some asymmetry, if this is sufficiently small there will be a wide temperature window in which it remains weak leading to small deviations from Eq.~\ref{ra}.

While present technology allows for spin polarized experiments, a useful sensor of local field
gradients must operate away from polarizing magnets. Therefore a spinful ladder must be considered; this case is more complex in details but conceptually contains similar physics, and will be addressed elsewhere \cite{dopo}.

It would be interesting to extend the present analysis to multi-wire
arrays, where it is expected that the single impurity should not
completely prohibit conductance as is the case in truly
two-dimensional systems.

{\em Acknowledgments -- } 
The authors thank the Abdus Salam ICTP for hospitality and D. Bagrets
and V.E. Kravtsov for insightful conversations.  A.A.N. is partly
supported by the grants GNSF-ST09/4-447 and SCOPES IZ73ZO-128058/1.


\begin{thebibliography}{10}
\bibitem{fn1} For inter-chain attraction $U<0$ ballistic
  conductance may occur only for an exotic case of the anti-symmetric
  external potential, i.e. when the two impurities on the rung are 
  of the same strength but opposite sign.
\bibitem{fn2} If $t_\perp=0$ but $U\ne 0$, one still observes a
  density modulation on the second chain. This modulation however is
  due to the inter-chain interaction. Thus the scattering problem
  is no longer of the lowest order in interaction and essentially is
  beyond the Hartree-Fock picture of Ref.~\onlinecite{mag}.
\end{thebibliography}
\end{document}